\documentclass[a4paper,11pt]{article}
\usepackage{amsmath,amssymb,amsfonts}
\usepackage[numbers,sort&compress,merge]{natbib}
\usepackage{graphicx}
\usepackage{rotating}
\usepackage[hang,bf,nooneline]{subfigure}
\usepackage[hang,bf,nooneline]{caption}
\newlength{\dslashwidth}

\newcommand{\bsg}{\ensuremath{b\to X_s\gamma}}

\newcommand{\tb}{\ensuremath{\tan\beta}}

\newcommand{\beq}{\begin{equation}} 
\newcommand{\eeq}{\end{equation}}
\newcommand{\beqa}{\begin{eqnarray}} 
\newcommand{\eeqa}{\end{eqnarray}}
\newcommand{\newc}{\newcommand}
\newcommand{\bq}{\begin{equation}}
\newcommand{\eq}{\end{equation}}
\newcommand{\ba}{\begin{array}}
\newcommand{\ea}{\end{array}}
\newcommand{\bqa}{\begin{eqnarray}}
\newcommand{\eqa}{\end{eqnarray}}

\newcommand{\lnf}{{\ifmmode \Lambda^{(N_f)} \else $\Lambda^{(N_f)}$\fi}}
\newcommand{\ms}{{\ifmmode \overline{MS} \else $\overline{MS}$\fi}}
\newcommand{\dr}{{\ifmmode \overline{DR} \else $\overline{DR}$\fi}}
\newcommand{\lms}{{\ifmmode \Lambda^{(5)}_{\overline{MS}} \else $\Lambda^{(5)}_{\overline{MS}}$\fi}}
\newcommand{\lam}{{\ifmmode \Lambda \else $\Lambda$\fi}}
\newcommand{\mev}{{\ifmmode {\rm MeV} \else ${\rm MeV}$\fi}}
\newcommand{\gev}{{\ifmmode {\rm GeV} \else ${\rm GeV}$\fi}}
\newcommand{\gevc}{{\ifmmode {\rm GeV/c^2} \else ${\rm GeV/c^2}$\fi}}
\newcommand{\tev}{{\ifmmode {\rm TeV} \else ${\rm TeV}$\fi}}
\newcommand{\tevc}{{\ifmmode {\rm TeV/c^2} \else ${\rm TeV/c^2}$\fi}}
\newcommand{\lp}{{\ifmmode L^+  \else $L^+$\fi}}
\newcommand{\lm}{{\ifmmode L^-  \else $L^-$\fi}}
\newcommand{\mlp}{{\ifmmode M(L^-) \else $M(L^-)$\fi}}
\newcommand{\mlz}{{\ifmmode M(L^0) \else $M(L^0)$\fi}}
\newcommand{\lz}{{\ifmmode L^0 \else $L^0$\fi}}
\newcommand{\ev}{{\ifmmode GeV/c^2 \else $GeV/c^2$\fi}}
\newcommand{\tri}{{\ifmmode \triangleup \else $\triangleup$\fi}}
\newcommand{\unl}{{\ifmmode U_{lL^0} \else $U_{lL^0}$\fi}}\newcommand{\gL}{{\ifmmode g_L \else $g_{L}$\fi}}
\newcommand{\gR}{{\ifmmode g_R  \else $g_{R}$\fi}}
\newcommand{\gumu}{{\ifmmode \gamma^{\mu} \else $\gamma^{\mu}$\fi}}
\newcommand{\gunu}{{\ifmmode \gamma^{\nu} \else $\gamma^{\nu}$\fi}}
\newcommand{\gdmu}{{\ifmmode \gamma_{\mu} \else $\gamma_{\mu}$\fi}}
\newcommand{\gdnu}{{\ifmmode \gamma_{\nu} \else $\gamma_{\nu}$\fi}}
\newcommand{\stw}{{\ifmmode\sin^2\theta_W \else $\sin^{2}\theta_{W}$ \fi}}
\newcommand{\sws}{{\ifmmode \;\sin^2\theta_W  \else $\;\sin^{2}\theta_{W}$ \fi}}
\newcommand{\cws}{{\ifmmode \;\cos^2\theta_W  \else $\;\cos^{2}\theta_{W}$ \fi}}
\newcommand{\sw}{{\ifmmode \;\sin\theta_W  \else $\sin\theta_{W}$ \fi}}
\newcommand{\cw}{{\ifmmode \;\cos\theta_W  \else $\;\cos\theta_{W}$ \fi}}
\newcommand{\tw}{{\ifmmode \;\tan\theta_W  \else $\;\tan\theta_{W}$ \fi}}
\newcommand{\qq}{{\ifmmode q\overline{q} \else $q\overline{q}$\fi}}
\newcommand{\lR}{{\ifmmode l_R \else $l_R$\fi}}
\newcommand{\lL}{{\ifmmode l_L \else $l_L$\fi}}
\newcommand{\nt}{{\ifmmode \nu_{\tau} \else $\nu_{\tau}$\fi}}
\newcommand{\nuR}{{\ifmmode \nu_R  \else $\nu_R$\fi}}
\newcommand{\nuL}{{\ifmmode \nu_L  \else $\nu_L$\fi}}
\newcommand{\qR}{{\ifmmode g_R  \else $q_R$\fi}}
\newcommand{\qL}{{\ifmmode q_L  \else $q_L$\fi}}
\newcommand{\qRp}{{\ifmmode q_R'  \else $q_{R}$'\fi}}
\newcommand{\qLp}{{\ifmmode q_L'  \else $q_{L}$'\fi}}
\newcommand{\est}{{\ifmmode e^{\bf \ast} \else $e^{\bf \ast}$\fi}}
\newcommand{\lst}{{\ifmmode l^{\bf \ast} \else $l^{\bf \ast}$\fi}}
\newcommand{\must}{{\ifmmode \mu^{\bf \ast} \else $\mu^{\bf \ast}$\fi}}
\newcommand{\taust}{{\ifmmode \tau^{\bf \ast} \else $\tau^{\bf \ast}$ \fi}}
\newcommand{\pperp}{{\ifmmode p_t  \else $p_t$\fi}}
\newcommand{\et}{{\ifmmode E_t  \else $E_t$\fi}}
\newcommand{\xt}{{\ifmmode x_t  \else $x_t$\fi}}
\newcommand{\smumu}{{\ifmmode \sigma_{\mu\mu}  \else $\sigma_{\mu\mu}$ \fi}}
\newcommand{\eg}{{\ifmmode e\gamma  \else $e\gamma$\fi}}
\newcommand{\epem}{{\ifmmode e^+e^-  \else $e^+e^-$\fi}}
\newcommand{\lplm}{{\ifmmode L^+L^-  \else $L^+L^-$\fi}}
\newcommand{\pp}{{\ifmmode p\overline p  \else $p\overline p$\fi}}
\newcommand{\llz}{{\ifmmode L^0\overline{L}^0 \else $L^0\overline{L}^0$\fi}}
\newcommand{\epemt}{{\ifmmode e^+e^- \to  \else $e^+e^- \to$\fi}}
\newcommand{\eb}{{\ifmmode E_{beam}  \else $E_{beam}$\fi}}
\newcommand{\ip}{{\ifmmode pb^{-1}  \else $pb^{-1}$\fi}}
\newcommand{\upm}{{\ifmmode ^{\pm}  \else $^{\pm}$\fi}}
\newcommand{\de}{{\ifmmode ^{\circ}  \else $^{\circ}$ \fi}}
\newcommand{\appr}{{\ifmmode \sim \else $\sim$ \fi}}
\newcommand{\corresp}{{\ifmmode \stackrel{\wedge}{=} \else $\stackrel{\wedge}{=}$ \fi}}
\newcommand{\sqrts}{{\ifmmode \sqrt{s} \else $\sqrt{s}$\fi}}
\newcommand{\zz}{{\ifmmode Z^0  \else $Z^0$\fi}}
\newcommand{\mz}{{\ifmmode M_{Z}  \else $M_{Z}$\fi}}
\newcommand{\mzs}{{\ifmmode M_{Z}^2  \else $M_{Z}^2$\fi}}
\newcommand{\mw}{{\ifmmode M_{W}  \else $M_{W}$\fi}}
\newcommand{\mws}{{\ifmmode M_{W}^2  \else $M_{W}^2$\fi}}
\newcommand{\mh}{{\ifmmode M_{Higgs}  \else $M_{Higgs}$\fi}}
\newcommand{\gt}{{\ifmmode \Gamma_{tot} \else $\Gamma_{tot}$\fi}}
\newcommand{\msusy}{{\ifmmode M_{SUSY}  \else $M_{SUSY}$\fi}}
\newcommand{\msusys}{{\ifmmode M_{SUSY}^2  \else $M_{SUSY}^2$\fi}}
\newcommand{\su}{{\ifmmode SU(3)_C\otimes\- SU(2)_L\otimes\- U(1)_Y \else $SU(3)_C\otimes SU(2)_L\otimes U(1)_Y$\fi}}
\newcommand{\suthree}{{\ifmmode SU(3)_C  \else $SU(3)_C$\fi}}
\newcommand{\sutwo}{{\ifmmode  SU(2)_L\otimes U(1)_Y \else $SU(2)_L\otimes U(1)_Y$\fi}}
\newcommand{\taup}{{\ifmmode \tau_{proton} \else $\tau_{proton}$\fi}}
\newcommand{\as}{{\ifmmode \alpha_{s}  \else $\alpha_{s}$\fi}}
\newcommand{\mgut}{{\ifmmode M_{GUT}  \else $M_{GUT}$\fi}}
\newcommand{\mguts}{{\ifmmode M_{GUT}^2  \else $M_{GUT}^2$\fi}}
\newcommand{\mzero}{{\ifmmode m_0        \else $m_0$\fi}}
\newcommand{\mhalf}{{\ifmmode m_{1/2}    \else $m_{1/2}$\fi}}
\newcommand{\sq}{{\ifmmode \tilde{q}    \else $\tilde{q}$\fi}}
\newcommand{\gl}{{\ifmmode \tilde{g}    \else $\tilde{g}$\fi}}
\newcommand{\mb}{{\ifmmode m_{b}    \else $m_{b}$\fi}}
\newcommand{\mt}{{\ifmmode m_{t}    \else $m_{t}$\fi}}
\newcommand{\mts}{{\ifmmode m_{t}^2    \else $m_{t}^2$\fi}}

\newcommand{\mtau}{{\ifmmode m_{\tau}  \else $m_{\tau}$\fi}}
\newcommand{\dpp}{{\ifmmode \delta_{pert} \else $\delta_{pert}$\fi}}
\newcommand{\dnp}{{\ifmmode\delta_{non-pert}\else$\delta_{non-pert}$\fi}}
\newcommand{\dew}{{\ifmmode \delta_{\rm EW}\else $\delta_{\rm EW}$\fi}}
\newcommand{\rt}{{\ifmmode R_{\tau}  \else $R_{\tau} $\fi}}
\newcommand{\rz}{{\ifmmode R_{Z}  \else $R_{Z} $\fi}}

\newcommand{\swb}{{\ifmmode \sin^2\theta_{\overline{MS}} \else $\sin^2\theta_{\overline{MS}}$\fi}}
\newcommand{\cwb}{{\ifmmode \cos^2\theta_{\overline{MS}} \else $\cos^2\theta_{\overline{MS}}$\fi}}

\newcommand{\bsmm}{\ensuremath{B^0_s\to\mu^+\mu^-}}

\newcommand{\btaunu}{\ensuremath{B_u\to\tau\nu}}

\newc\AIPCP[3] {{\em AIP Conf. Proc.} {\bf #1} (#2) #3}
\newc\AJ[3] {{\em Astrophys. J.} {\bf #1} (#2) #3}
\newc\AMS[3] {{\em Ann. Math. Statist.} {\bf #1} (#2) #3}
\newc\AP[3] {{\em Ann. Phys.} {\bf #1} (#2) #3}
\newc\APJ[3] {{\em Astropart. J.} {\bf #1} (#2) #3}
\newc\APP[3] {{\em Astropart. Phys.} {\bf #1} (#2) #3}
\newc\APS[3] {{\em Astrophys. J. Suppl.} {\bf #1} (#2) #3}
\newc\ARNPS[3] {{\em Ann. Rev. Nucl. Part. Sci.} {\bf C#1} (#2) #3}
\newc\BA[3] {{\em Bayesian Anal.} {\bf C#1} (#2) #3}
\newc\CPC[3] {{\em Comput. Phys. Commun.} {\bf C#1} (#2) #3}
\newc\CP[3] {{\em Contemp. Phys.} {\bf #1} (#2) #3}
\newc\EPJ[3] {{\em Euro. Phys. Journ.} {\bf C#1} (#2) #3}
\newc\JCAP[3] {{\em JCAP} {\bf #1} (#2) #3}
\newc\JHEP[3] {{\em JHEP} {\bf #1} (#2) #3}
\newc\JPG[3] {{\em J. Phys.} {\bf G #1} (#2) #3}
\newc\IJMP[3] {{\em Int. J. Mod. Phys.} {\bf A #1} (#2) #3}
\newc\MNRAS[3] {{\em Mon. Not. Roy. Astron. Soc.} {\bf #1} (#2) #3}
\newc\MPL[3] {{\em Mod. Phys. Lett.} {\bf A #1} (#2) #3}
\newc\NAR[3] {{\em New Astron. Rev.} {\bf #1} (#2) #3}
\newc\NCA[3] {{\em Nuovo Cimento} {\bf #1} (#2) #3}
\newc\NIM[3] {{\em Nucl. Instrum. Methods} {\bf #1} (#2) #3}
\newc\NIMA[3] {{\em Nucl. Instrum. Methods} {\bf A #1} (#2) #3}
\newc\NAT[3] {{\em Nature} {\bf #1} (#2) #3}
\newc\NPB[3] {{\em Nucl. Phys.} {\bf B #1} (#2) #3}
\newc\NPA[3] {{\em Nucl. Phys.} {\bf A #1} (#2) #3}
\newc\NPPS[3] {{\em Nucl. Phys. Proc. Suppl.} {\bf #1} (#2) #3}
\newc\PLB[3] {{\em Phys. Lett.} {\bf B #1} (#2) #3}
\newc\PR[3] {{\em Phys. Rep.} {\bf #1} (#2) #3}
\newc\PRL[3] {{\em Phys. Rev. Lett.} {\bf #1} (#2) #3}
\newc\PRD[3] {{\em Phys. Rev.} {\bf D #1} (#2) #3}
\newc\PRC[3] {{\em Phys. Rev.} {\bf C #1} (#2) #3}
\newc\PTP[3] {{\em Prog. Theor. Phys.} {\bf #1} (#2) #3}
\newc\RMP[3] {{\em Rev. Mod. Phys.} {\bf #1} (#2) #3 }
\newc\RPP[3] {{\em Rept. Prog. Phys.} {\bf #1} (#2) #3 }
\newc\SC[3] {{\em Science} {\bf #1} (#2) #3 }
\newc\ZPC[3] {{\em Z. Phys.} {\bf C #1} (#2) #3}
\newc\Err[3] {{\em Erratum-ibid.} {\bf #1} (#2) #3 }

\begin{document}
\begin{center}

\Large \textbf{Constraints on Supersymmetry from LHC data on SUSY searches and Higgs bosons combined with cosmology and direct dark matter searches}
\vspace{10mm}

\large

C. Beskidt$^a$, W. de Boer$^{a}$\footnote{Email: wim.de.boer@kit.edu}, D.I. Kazakov$^{b,c}$, F. Ratnikov$^{a,c}$

\normalsize
\vspace{5mm}
$^a$ \textit{Institut f\"ur Experimentelle Kernphysik,
Karlsruhe Institute of Technology,\\ P.O. Box 6980, 76128 Karlsruhe, Germany}

\vspace{5mm}
$^b$ \textit{Bogoliubov Laboratory of Theoretical Physics, Joint Institute for Nuclear Research,\\
141980, 6 Joliot-Curie, Dubna, Moscow Region, Russia}

\vspace{5mm}
$^c$ \textit{Institute for Theoretical and Experimental Physics,\\
117218, 25 B.Cheremushkinskaya, Moscow, Russia}

\vspace{30mm} \textbf{Abstract} \vspace{5mm}

\begin{minipage}[c]{12cm}

\textit{
The ATLAS and CMS experiments did not find evidence for Supersymmetry using close to 5/fb of published LHC data at a center-of-mass energy of 7 TeV. We combine these LHC data with data on $\bsmm$ (LHCb experiment), the relic density (WMAP and other cosmological data) and upper limits on the dark matter scattering cross sections on nuclei (XENON100 data). The excluded regions in the constrained Minimal Supersymmetric SM (CMSSM) lead to gluinos excluded below 1270 GeV and dark matter candidates below 220 GeV for values of the scalar masses ($m_0$) below 1500 GeV. For large $m_0$ values the limits of the gluinos and the dark matter candidate are reduced to 970 GeV and 130 GeV, respectively.
If a Higgs mass of 125 GeV is imposed in the fit, the preferred SUSY region is above this excluded region, but
the size of the  preferred region is strongly dependent on the assumed theoretical error.
}
\end{minipage}

\end{center}

\thispagestyle{empty}
\setcounter{page}{0}

\section{Introduction}
Supersymmetry (SUSY) is a good candidate for physics beyond the SM, because it solves the hierarchy problem, allows for unification of the coupling constants and predicts electroweak symmetry breaking (EWSB) with the lightest Higgs mass below 130 GeV (for reviews, see e.g. \cite{Haber:1984rc,deBoer:1994dg,Martin:1997ns,Kazakov:2010qn}). In addition the Lightest Supersymmetric Particle (LSP) has all the properties expected for the Weakly Interacting Massive Particles (WIMPS) of the dark matter \cite{Kolb:1990vq,Jungman:1995df,Bertone:2004pz}, which is known to make up more than 80\% of the matter in the universe \cite{Komatsu:2010fb}. From the observed DM density, which is proportional to the inverse of the annihilation cross section one finds a strong constraint, if one assumes all DM is made from supersymmetric LSPs.
Unfortunately, direct searches for the predicted SUSY particles at the LHC running at 7 TeV were unsuccessful. Also direct DM searches in deep underground experiments were  contradictory \cite{Bertone:2004pz}. 
Combining all data from the LHC, cosmology and direct DM searches leads to  strong constraints on the masses of the predicted SUSY masses, as discussed in many recent papers \cite{Buchmueller:2011ki,Buchmueller:2011sw,Bertone:2011nj,Allanach:2011ut,
Allanach:2011wi,Farina:2011bh,Strumia:2011dy,Akula:2011dd,Trotta:2008bp,Akrami:2009hp,Feroz:2008wr,Sekmen:2011cz}.

To restrict the number of independent SUSY masses one usually assumes the masses to be unified at the GUT scale
and the particles get different masses at lower energies because of radiative corrections. This works well for the SM model particles of the third generation, if they are in the same multiplet and get mass from the same Higgs field. E.g. the 
ratio of $b/\tau$ masses is well predicted by radiative corrections, if one assumes the Yukawa couplings are unified at the GUT scale. For the mass breaking terms of the SUSY particles
one assumes that the masses of spin 0 (spin 1/2) particles are unified at the GUT scale with values $\mzero (\mhalf)$.
In the constrained Minimal Supersymmetric SM (CMSSM) \cite{Chamseddine:1982jx,Kolda:1994ab} the many parameters of SUSY models  are reduced to only four: the two mass parameters $\mzero$, $\mhalf$ and two parameters related to the Higgs sector: the trilinear coupling at the GUT scale $A_0$, and \tb, the ratio of the vacuum expectation values of the two neutral components of the two Higgs doublets. Electroweak symmetry breaking (EWSB) fixes the scale of $\mu$, so only its sign is a free parameter.
  The positive sign is taken, as
suggested by the small deviation of the SM prediction from the muon anomalous moment, see e.g. \cite{deBoer:2001nu}.

In this letter we combine the newest data from LHC, WMAP, XENON100, flavor physics and g-2. The specific observables are detailed in Table \ref{t1}. 
We start by discussing the fitting technique, the observations and the excluded regions of each observation separately. The combination of all constraints differs from the results from other groups, which is attributed to the fitting technique.

\section{Multistep Fitting Technique}\label{multi}
Excluded regions have been determined by many different groups  either using a frequentist approach by maximizing a likelihood  or using random sampling techniques of the parameter space, see e.g. \cite{Buchmueller:2011ki,Buchmueller:2011sw,Bertone:2011nj,Allanach:2011ut,
Allanach:2011wi,Farina:2011bh,Strumia:2011dy,Akula:2011dd,Trotta:2008bp,Akrami:2009hp,Feroz:2008wr,Sekmen:2011cz} and references therein. Bayesian techniques, as typically used with Markov Chain or Multinest sampling techniques, are dependent on the prior, which leads to an additional, non-quantifiable uncertainty in the excluded or allowed regions, see e.g. \cite{Feroz:2011bj} for a recent discussion and references therein.  We believe this uncertainty is due to the high correlations between three of the four parameters, as we discussed in two previous papers \cite{Beskidt:2011qf,Beskidt:2010va}.
Such strong correlations lead to
likelihood spikes in the parameter region, where three of the four parameters have to have specific correlated values. Although the likelihood of such narrow regions is high, they can be easily missed in methods based on stepping techniques.

\begin{table}
\centering
\begin{tabular}{lll}
\hline\noalign{\smallskip}
Constraint & Data & Ref.  \\
\noalign{\smallskip}\hline\noalign{\smallskip}
$\Omega h^2$ & $0.113\pm 0.004$ & \cite{Komatsu:2010fb} \\
 $\bsg$ & $(3.55 \pm 0.24)\cdot 10^{-4}$ & \cite{hfag} \\
   $\btaunu$ &  $(1.68\pm 0.31)\cdot 10^{-4}$ & \cite{hfag} \\
    $\Delta a_\mu$ & $(302~\pm~63 (exp)~\pm~61 (theo))\cdot 10^{-11}$ &  \cite{Bennett:2006fi}\\
   $\bsmm$ &  $\bsmm < 4.5\cdot 10^{-9}$ & \cite{Aaij:2012ac}\\
$m_h$  & $ m_h > 114.4$ GeV & \cite{Schael:2006cr}\\
$m_A$ & $m_A > 480$ GeV for $\tb \approx 50$& \cite{Chatrchyan:2012vp,Aad:2011rv}\\
ATLAS & $ \sigma^{SUSY}_{had} < 0.003-0.03 $ pb & \cite{ATLAS-CONF-2012-033}\\
CMS & $\sigma^{SUSY}_{had} < 0.005-0.03 $ pb &\cite{CMS-PAS-SUS-12-005} \\
XENON100 & $\sigma_{\chi N} < 8 \cdot 10^{-45}-2\cdot 10^{-44} cm^2$& \cite{Aprile:2011hi}\\
\noalign{\smallskip}\hline
\end{tabular}
\caption{List of all constraints used in the fit to determine the excluded region of the CMSSM parameter space. }
\label{t1}
\end{table}

To cope with the strong correlations we use a multistep fitting technique, defined by fitting the parameters with the strongest correlation first, i.e. we fit first $\tb$ and $A_0$ for each pair of the mass parameters  $\mzero$ and $\mhalf$ by  minimizing the $\chi^2$ with the program Minuit \cite{James:1975dr}. 
The most probable region of parameter space is determined by the minimum $\chi^2_{min}$ value. For two degrees of freedom the given 95\% C.L.  (90\% C.L.) limit is reached for an increase in $\chi^2$ of 5.99 (4.61).  The $\chi^2$ function is defined as 
\begin{equation}\label{chi_def}
\begin{split}
\chi^2=\frac{\left ( \Omega h^2 - 0.1131 \right)^2 }{\sigma^2_{\Omega h^2}}+ \frac{\left ( \bsg - 3.55\cdot 10^{-4} \right)^2 }{\sigma^2_{\bsg}}\\
+ \frac{\left ( \btaunu - 1.68 \cdot 10^{-4} \right)^2 }{\sigma^2_{\btaunu}} + \frac{\left ( \Delta a_\mu - 302~\cdot 10^{-11} \right)^2 }{\sigma^2_{\Delta a_\mu}}\\ 
+ \chi^2_{\bsmm} +\chi^2_{m_h}+\chi^2_{CMS}+\chi^2_{ATLAS}+ \chi^2_{m_A}+\chi^2_{DDMS}
\end{split} 
\end{equation}

For the first four terms in eq.~\ref{chi_def} the $\chi^2$ is defined in a straightforward way: the square of the difference between the predicted value for given SUSY parameters and the experimental value, weighted by the inverse of the error squared. For the remaining terms only 95\% C.L. or 90\% C.L. limits have been given.  All experimental values and limits have been summarized in Table~\ref{t1}. We discuss the remaining contributions to the $\chi^2$  separately:

\begin{itemize}
\item $\chi^2_{\bsmm}=\left( \bsmm - \bsmm_{95} \right)^2/\sigma^2_B$

It is added only if $\bsmm  > \bsmm_{95}$. $\bsmm_{95}$ was set such that with the theoretical error $\sigma_B = 0.2 \cdot 10^{-9}$  \cite{Aaij:2012ac} a change in $\chi^2$ of 5.99 was reached for the experimental 95\% C.L. upper limit from Table \ref{t1}. This leads to $\bsmm_{95}=4.0 \cdot 10^{-9}$.
 
\item $\chi^2_{m_h}=\left( m_h - m_h^{95} \right)^2/\sigma^2_h$

This contribution takes the lower limit on the Higgs mass into account. It is added to the $\chi^2$ function if $m_h < m_h^{95}$. Here a 95\% C.L. lower limit of $m_h$ of 114.4 GeV is used, as determined from the LEP experiments. The experimental error taken from the 1$\sigma$ band from Ref. \cite{Schael:2006cr} corresponds to the chosen $\sigma_h$ of 0.5 GeV, which result in $m^{95}_h=115.6$ GeV from the requirement $\Delta\chi^2=$5.99 for $m_h > $ 114.4 GeV at 95\% C.L..  

\item $\chi^2_{CMS}$ and  $\chi^2_{ATLAS}$

 These contributions take into account the negative results from the direct searches for SUSY particles from CMS and ATLAS, respectively. Low SUSY masses are excluded by these searches, since these would have a too large cross section. From the published 95\% C.L. exclusion contours in the ($\mzero,\mhalf$)-plane \cite{ATLAS-CONF-2012-033,CMS-PAS-SUS-12-005} one can determine the excluded SUSY cross sections, which vary  along the contour presumably because of the varying efficiency. Since the efficiency is not published, we take the following simplified approach for the $\chi^2$ contribution: we know that at the contour $\Delta \chi^2$=5.99 and we assume that the exclusion limit near the contour is proportional to the SUSY hadronic cross section $\sigma_{tot} \left ( pp \rightarrow \tilde{g}\tilde{g},\tilde{g}\tilde{q},\tilde{q}\tilde{q} \right )$, which is a reasonable approximation for the LHC limits as will be shown later in section~\ref{lhc}. Hence we parametrize $\chi^2_{LHC}=\sigma^2_{tot} / \sigma^2_{LHC}$, where LHC stands either for CMS or ATLAS and for each experiment $ \sigma^2_{LHC}$ was determined as function of  $\mzero$ and $\mhalf$ by the requirement that $\Delta \chi^2$=5.99 at the 95\% C.L. contour. 

\item $\chi^2_{m_A}$

This contribution takes care of the results of the search for the neutral Higgs bosons decaying into tau pairs in pp collisions at the LHC. The 95\% C.L. exclusion curve in the $\tb-m_A$ plane excludes for a given value of $\tb$ a certain mass of the pseudo-scalar Higgs  \cite{Chatrchyan:2012vp,Aad:2011rv}. 
The $\chi^2$ contour is calculated as $\chi^2=(m_A^{th}(\tb)-m_A^{95}(\tb))^2/\sigma^2_{m_A^{95}}$, where $\sigma_{m_A^{95}}$ can be obtained from the $1\sigma$ curves around the 95\% C.L. contour line. $m_A^{95}(\tb)$ is determined by the requirement that $\chi^2_{m_A}$ is 5.99, for $m_A^{th}(\tb)$ on the contour line.  

\item $\chi^2_{\chi N}$

This term takes the upper limits on the elastic WIMP-nucleon cross section $\sigma_{\chi N}$ of the direct dark matter searches into account. Usually only  90\% C.L. upper limit on $\sigma_{\chi N}$  are  given as function of the  WIMP mass $m_{WIMP}$ \cite{Aprile:2011hi}. The $\chi^2$ contribution can be included  in a  way similar to $\chi^2_{m_A}$.

 We define $\chi^2_{\chi N}=(\sigma_{\chi N}^{th}(m_{\chi})-\sigma_{\chi N}^{95}(m_{\chi}))^2/\sigma^2_{\chi N}(m_{\chi})$. The weight 1/$\sigma^2_{\chi N}$ is taken from the 1$\sigma$ band in Ref. \cite{Aprile:2011hi}. The excluded cross section $\sigma_{\chi N}^{95}(m_{\chi})$ is determined by requiring $\chi^2_{\chi N}=$4.61, since the published limit is at 90\% C.L.. 
\end{itemize}

All observables  were calculated with the public code micrOMEGAs 2.4.5
\cite{Belanger:2010pz,Pukhov:2010px} combined with Suspect 2.41 as  mass
spectrum calculator \cite{Djouadi:2002ze}. 
Within our multistep fitting technique we minimize the $\chi^2$ function defined in eq.~\ref{chi_def}. With the multiple minimization techniques inside Minuit and starting with only two parameters for each point in the ($\mzero$,$\mhalf$)-plane the program is fast and quickly converges to the global minimum.
Once the SUSY parameters have been fitted, one should also vary the SM parameters or marginalize over them, like the top and bottom mass and the strong coupling constant. The SM parameters are given in the Particle Data Book \cite{Nakamura:2010zzi}: we use $m_{top}^{pole}=172.5\pm 1.3$ and $m_b(m_b)^{\overline{MS}}=4.25\pm 0.2$ GeV for the heavy quark  masses  and  $\alpha_s=0.1172\pm 0.02$ for the strong coupling constant. However, the SM quark masses mainly determine the running of the Higgs mass parameters, so different values of these masses can be  compensated by a slightly different values of the SUSY parameters $A_0$ and $\tb$ in order to get the same $\chi^2$ value, so the allowed region is hardly affected.

\section{Combination of all Constraints}
\begin{figure} 
 \begin{center}
 \includegraphics[width=0.49\textwidth]{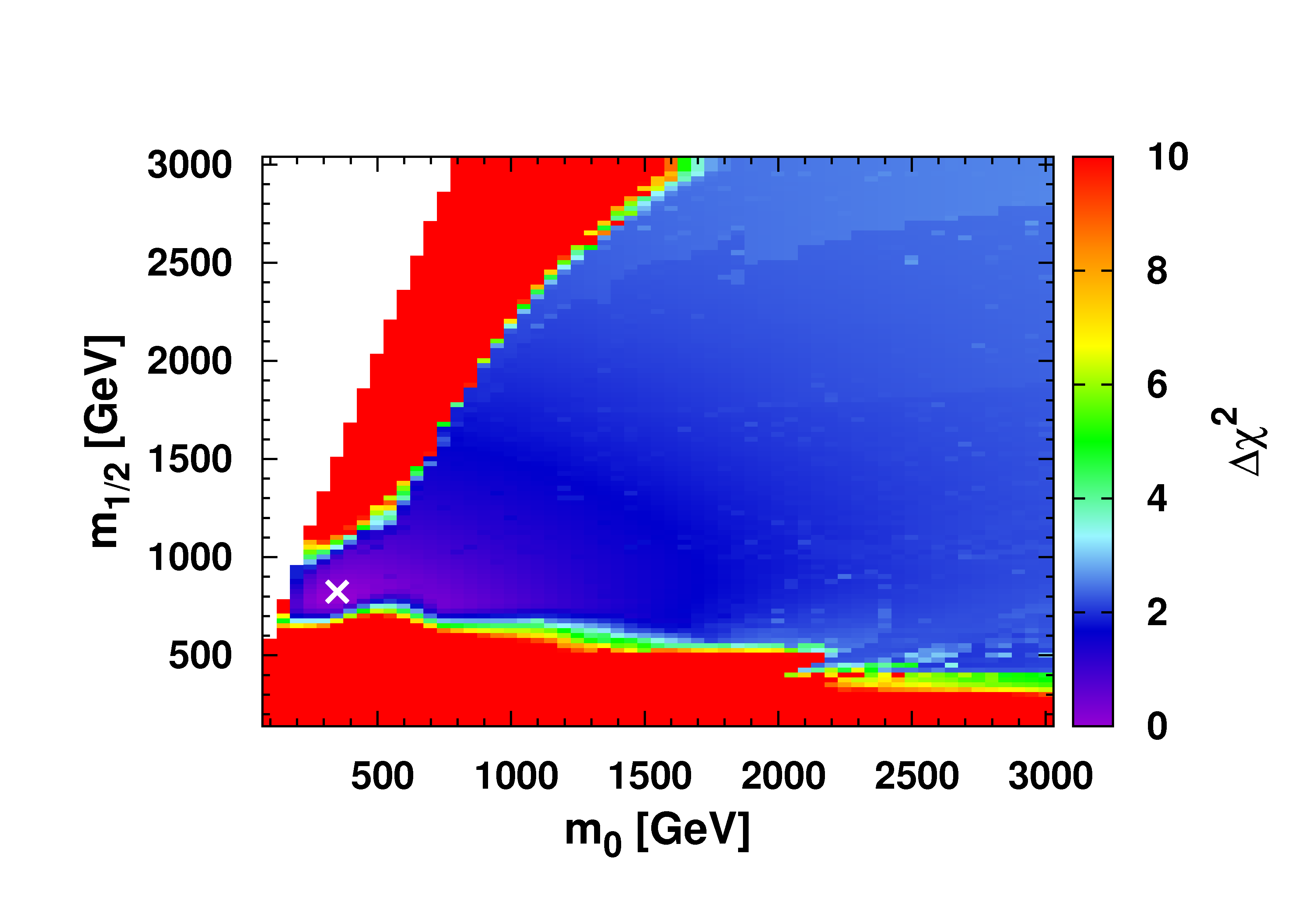}
 \includegraphics[width=0.49\textwidth]{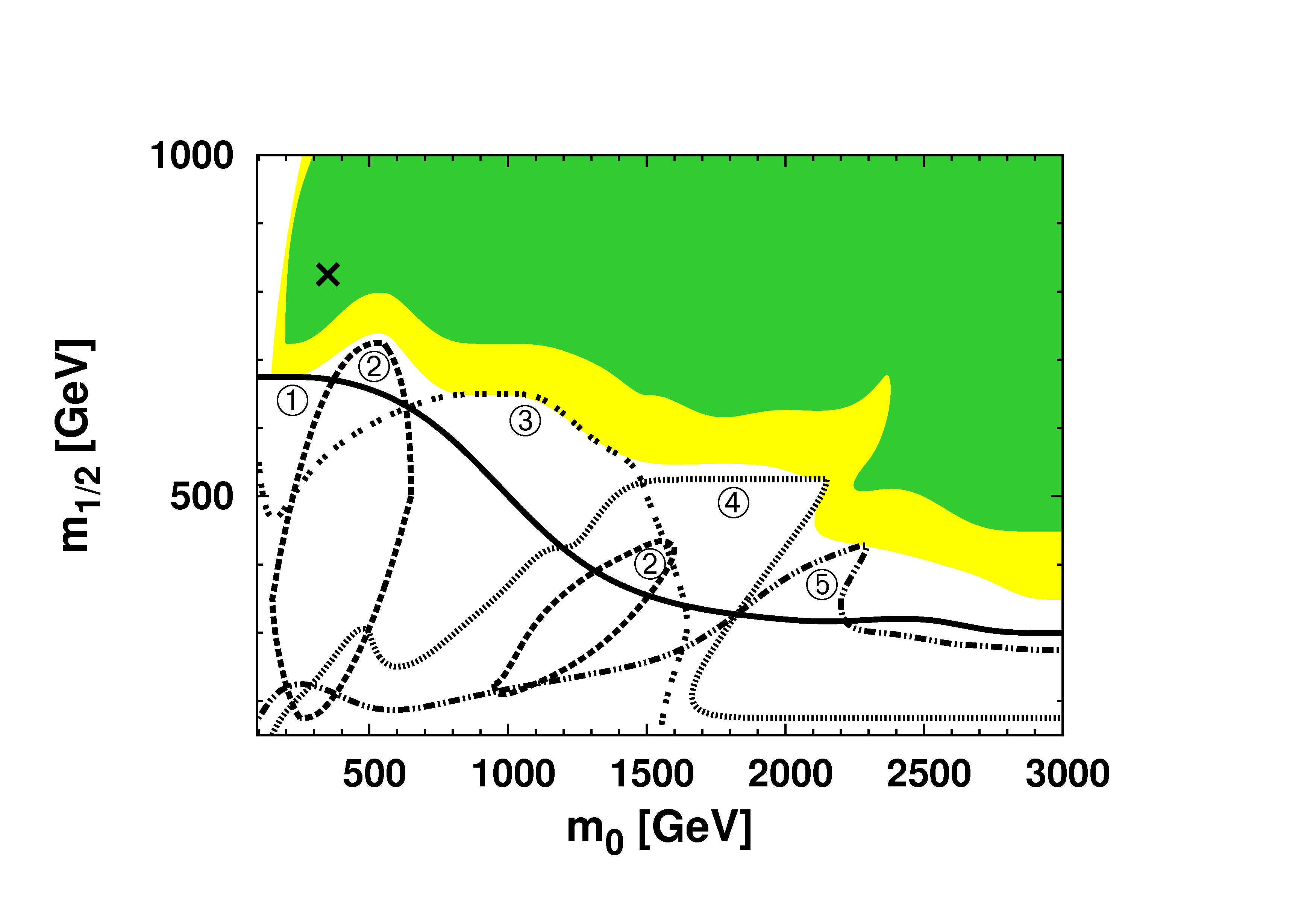} 
 \end{center}
 \caption{Left: $\Delta\chi^2$ distribution of all constraints up to $\mhalf=3000$ GeV. The white cross represents $\chi^2_{min}=4.06$. The white region in the top left corner is excluded because
 the stau is the LSP. The red region in this corner is excluded by the relic
 density constraint requiring large $\tb$, which in turn causes a large mixing in the stau sector leading
 to the  stau becoming the LSP again. Right: contributions to the $\chi^2$ of all constraints up to $\mhalf=1000$ GeV. The contour for each constraint represents the 95\% C.L. exclusion limit ($\Delta\chi^2=5.99$) for each constraint separately.}\label{f1}
 \end{figure}

\begin{figure}
 \begin{center}
 \includegraphics[width=0.49\textwidth]{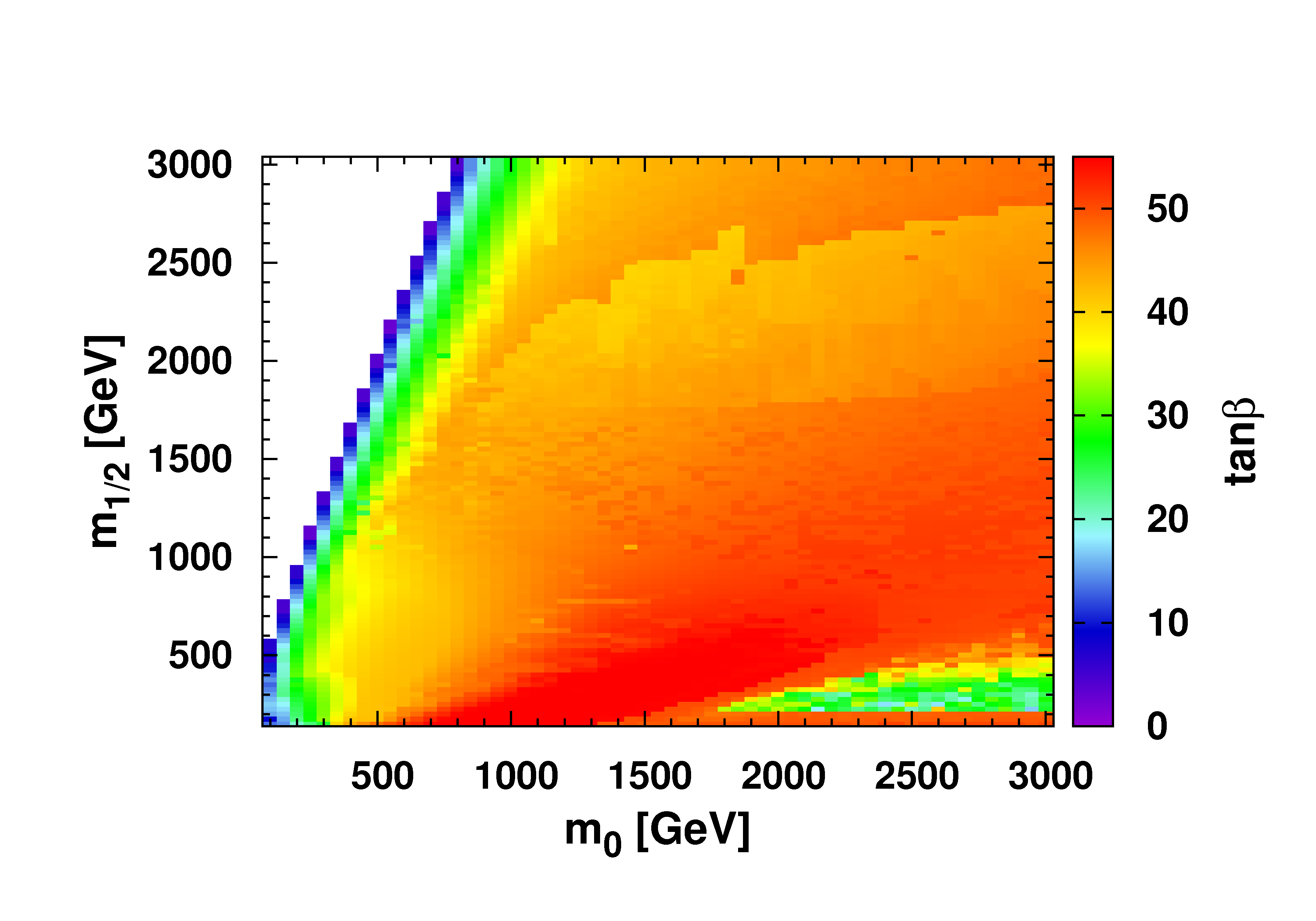}
 \includegraphics[width=0.49\textwidth]{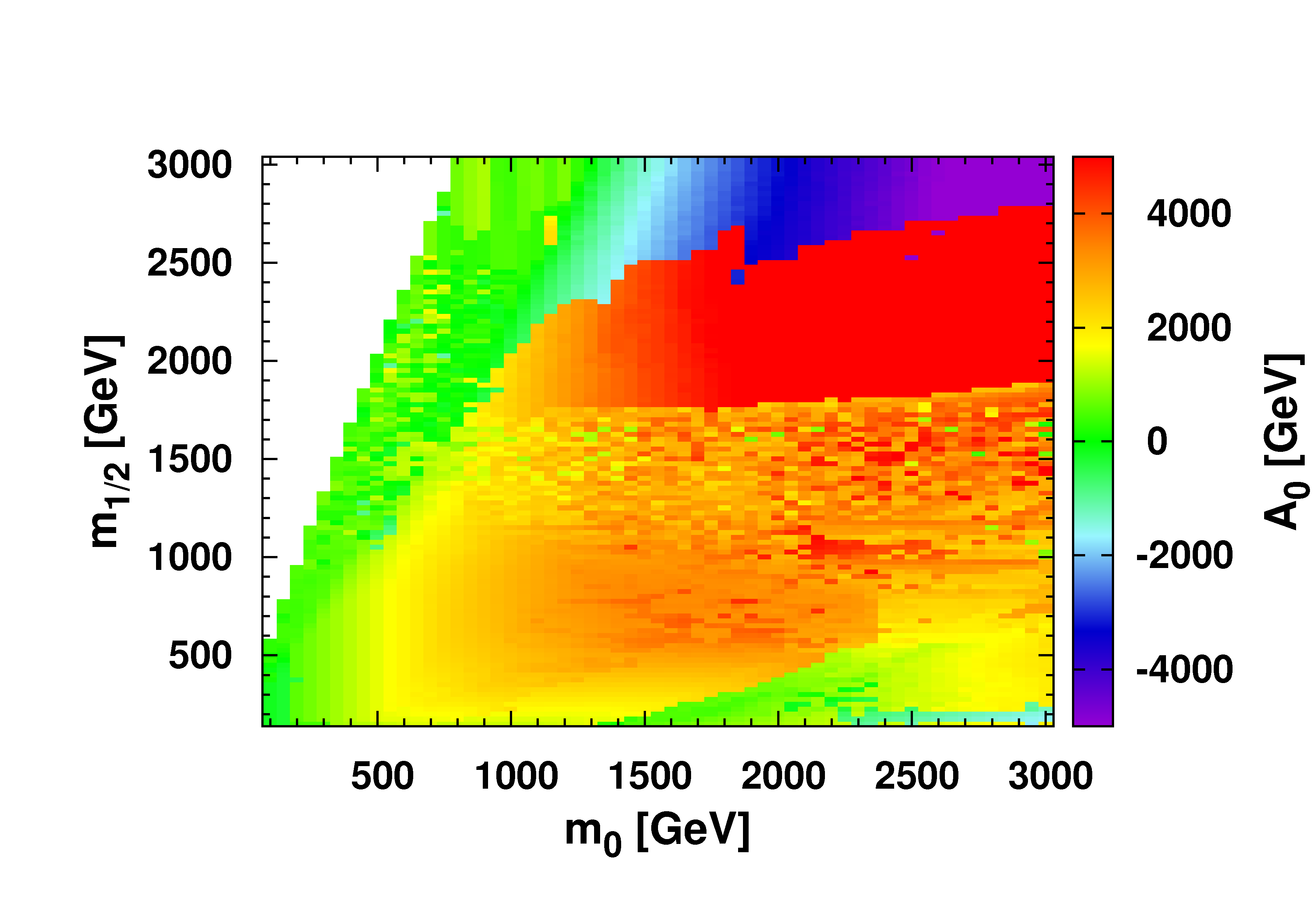}
 \end{center}
 \caption{Resulting $\tb$ (left) and $A_0$ (right) values from the fit to all data. }\label{f2}
 \end{figure}

The minimal values of the $\chi^2$ function of Eq. \ref{chi_def} in the ($\mzero,\mhalf$)-plane  are shown in the left panel of Fig. \ref{f1} for each $m_0,m_{1/2}$ pair. The color coding indicates the  value of  $\Delta\chi^2= \chi^2-\chi^2_{min}$, where the  the minimal value $\chi^2_{min}=4.06$ is obtained for the mass values $\mzero=350$ GeV, $\mhalf=825$ GeV, as indicated by the white cross.  The red and yellow regions, corresponding to $\Delta\chi^2>5.99$, are excluded at 95\% C.L.. The observables contributing most strongly to the exclusion vary in the plane as indicated in the right panel of Fig. \ref{f1}. These contours are drawn in the following way: we take $\Delta\chi_i^2=\chi^2_i-\chi^2_{i,min}=5.99$, where $\chi^2_{i,min}$ is the $\chi^2$-contribution of variable i at the best-fit point and $\chi^2_i$ is the $\chi^2$ value of variable i at the contour.
The direct SUSY searches at the LHC (contour 1) dominate the limit at small $m_0$ with a small contribution from the branching ratio of $\bsmm$ (contour 2). Other contributions at intermediate SUSY masses come from the Higgs searches (contour 3 for the SM Higgs and contour 4 for the pseudo-scalar Higgs) and direct DM searches (contour 5) at larger values of $m_0$. 
The fitted values of $\tb$ and $A_0$ for each $\mzero,\mhalf$ pair are shown in Fig.~\ref{f2}.

To understand the  contours in the right panel of Fig. \ref{f1}, we discuss each of them in more detail after discussing the influence of g-2 first.

\subsection{Influence of the anomalous magnetic moment of the muon}
\begin{figure}
 \begin{center}
 \includegraphics[width=0.49\textwidth]{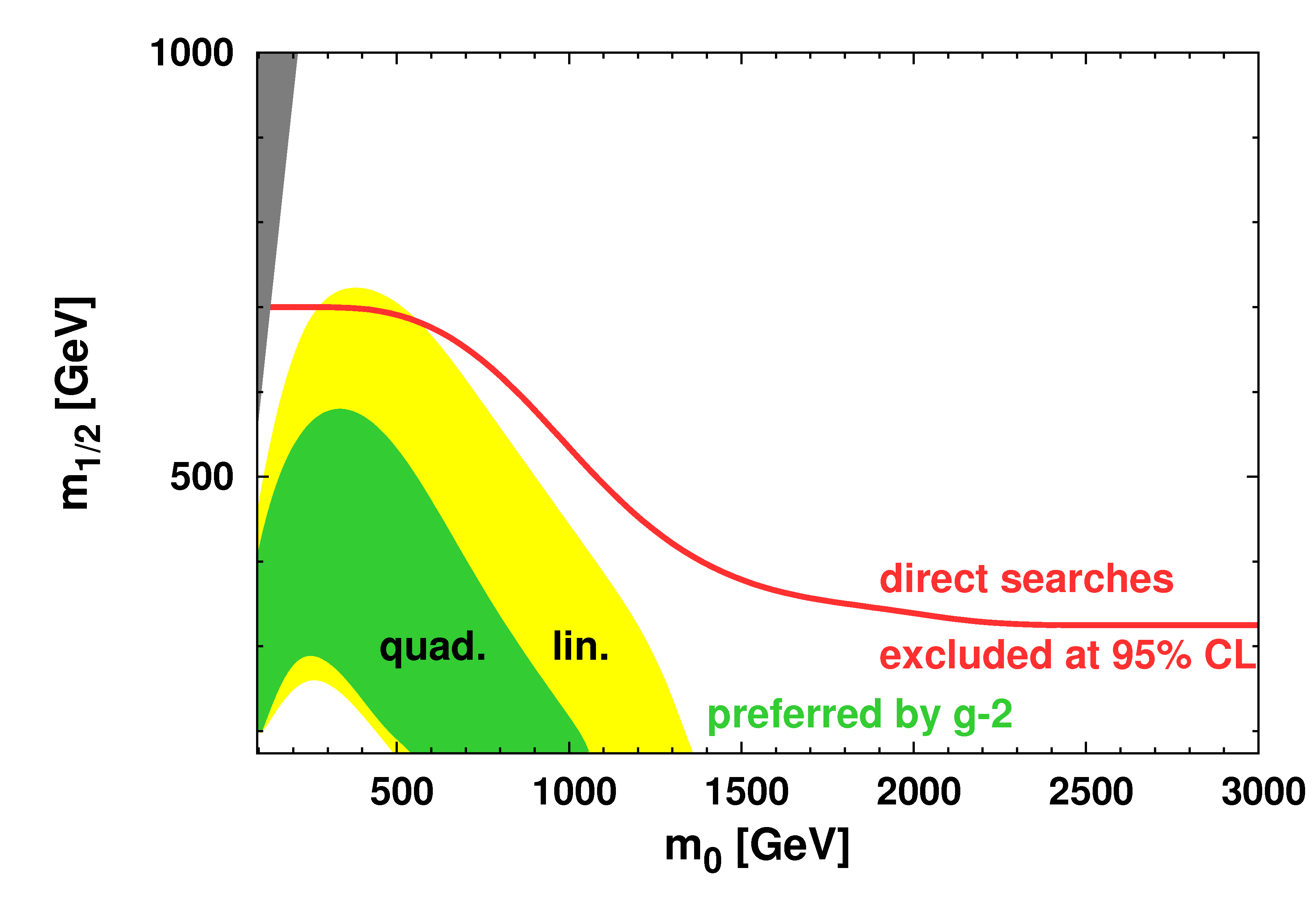}
 \end{center}
 \caption{ Preferred region of the g-2 observable alone under 
 the constraint that \tb~ and $A_0$ are fixed by all other constraints. Here we show the 1$\sigma$ band $(\Delta\chi^2 = 2.3)$ of the preferred region for quadratic (green, dark shaded) and linear (yellow, light shaded) addition of the theoretical and experimental errors. We compare these bands with the 68\% C.L. exclusion limit of the direct searches at the LHC. The preferred region by g-2 is largely excluded by the LHC constraints.  }\label{f3}
 \end{figure}

We included the value of g-2 into the fit, but as can be seen from Fig. \ref{f3} most of the preferred region by the g-2 constraint is  excluded by the direct searches of the LHC independent of the treatment of the g-2 uncertainties. The green region (dark shaded) is preferred by g-2 data if one adds the experimental and theoretical errors in quadrature. However, since these errors are of the same order of magnitude (see Table. \ref{t1}) and the theoretical uncertainties are certainly non-gaussian, a linear addition of the errors is more conservative, which leads to a larger error and hence a larger (yellow, light shaded) preferred region. However, even this larger region is still excluded by the LHC, so the observed three sigma deviation of the anomalous magnetic moment of the muon above the SM prediction may either be a statistical fluctuation or has an origin different from SUSY, if we assume the theoretical and experimental errors have been estimated correctly.

 \begin{figure} 
 \begin{center}
 \includegraphics[width=0.49\textwidth]{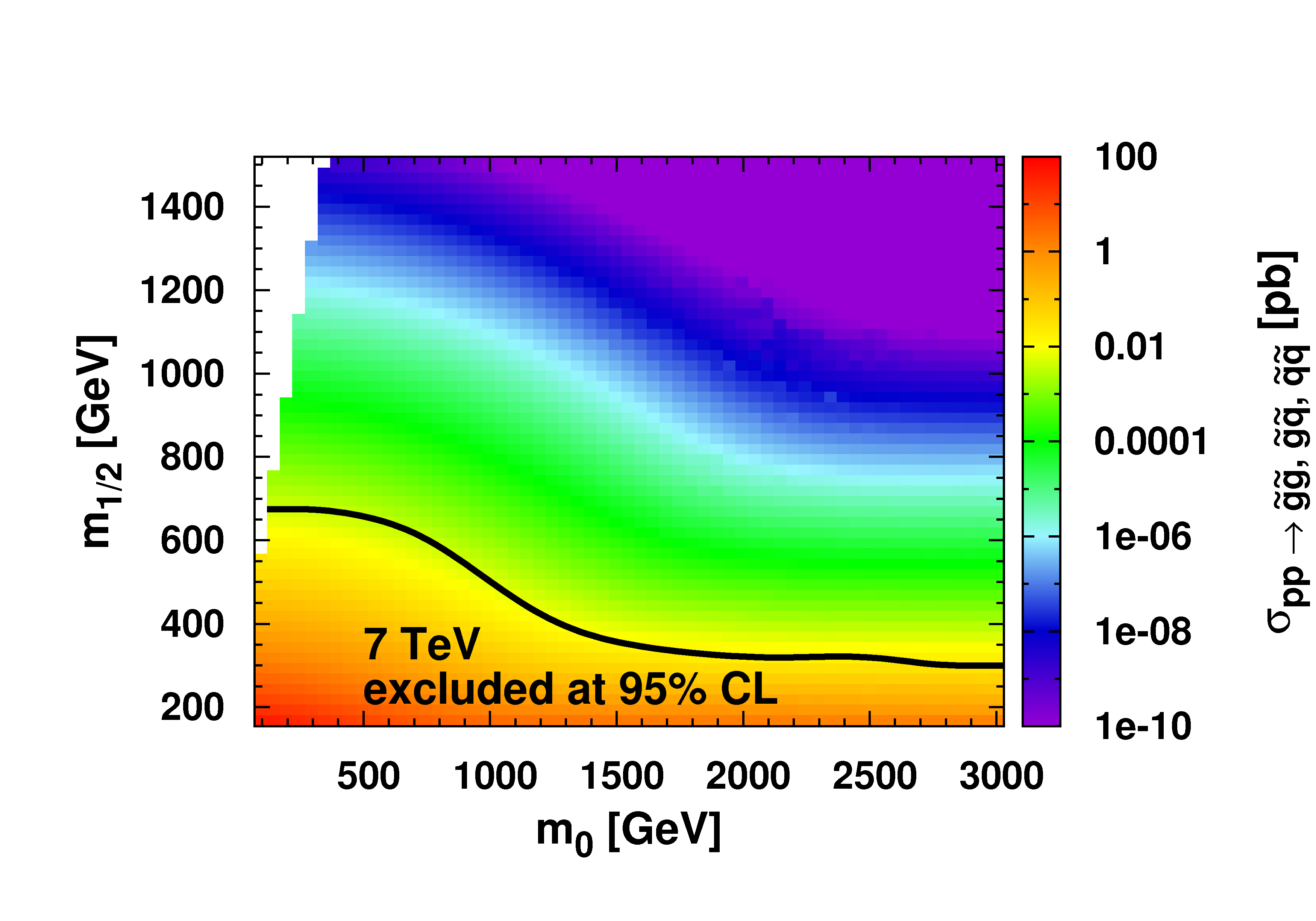}
\includegraphics[width=0.49\textwidth]{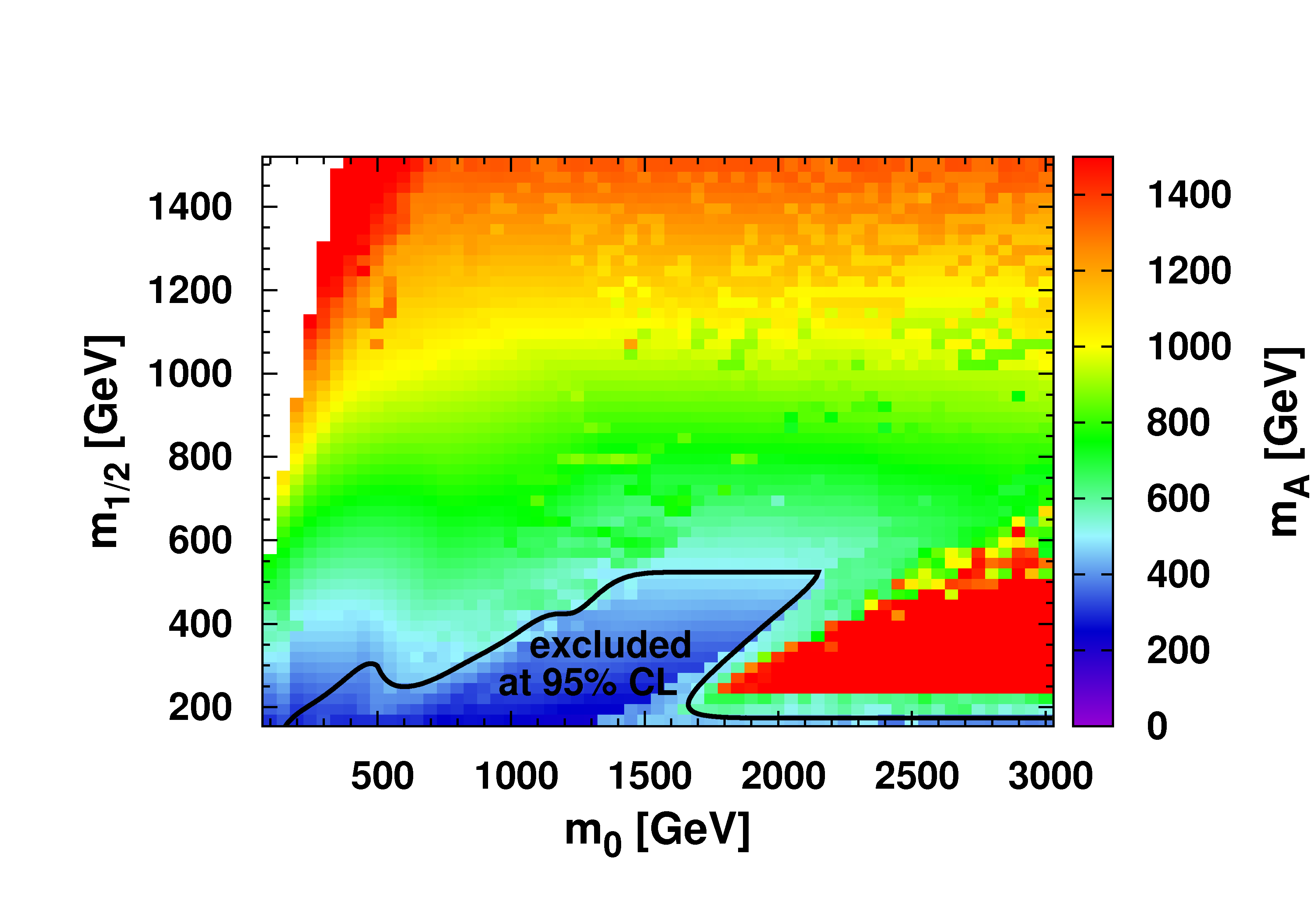}
 \end{center}
 \caption{Left: Total production cross section of strongly interacting particles (color coding) in comparison with the LHC excluded limits for 7 TeV. Here the data from ATLAS and CMS were combined. The ATLAS and CMS data correspond to an integrated luminosity of 4.4 and 4.71 fb$^{-1}$, respectively. One observes from the color coding that a cross section of 0.003 to 0.03 pb is excluded at 95\% confidence level. Right: Values of $m_A$ in the ($\mzero,\mhalf$)-plane after optimizing
 \tb~and $A_0$ to fulfill all constraints at every point. The data below the solid line in the right panel are excluded at 95\% confidence level from the $m_A$ exclusion contour  as function of $\tb$.  }
 \label{f4}
 \end{figure} 
 
\begin{figure} 
 \begin{center}
 \includegraphics[width=0.49\textwidth]{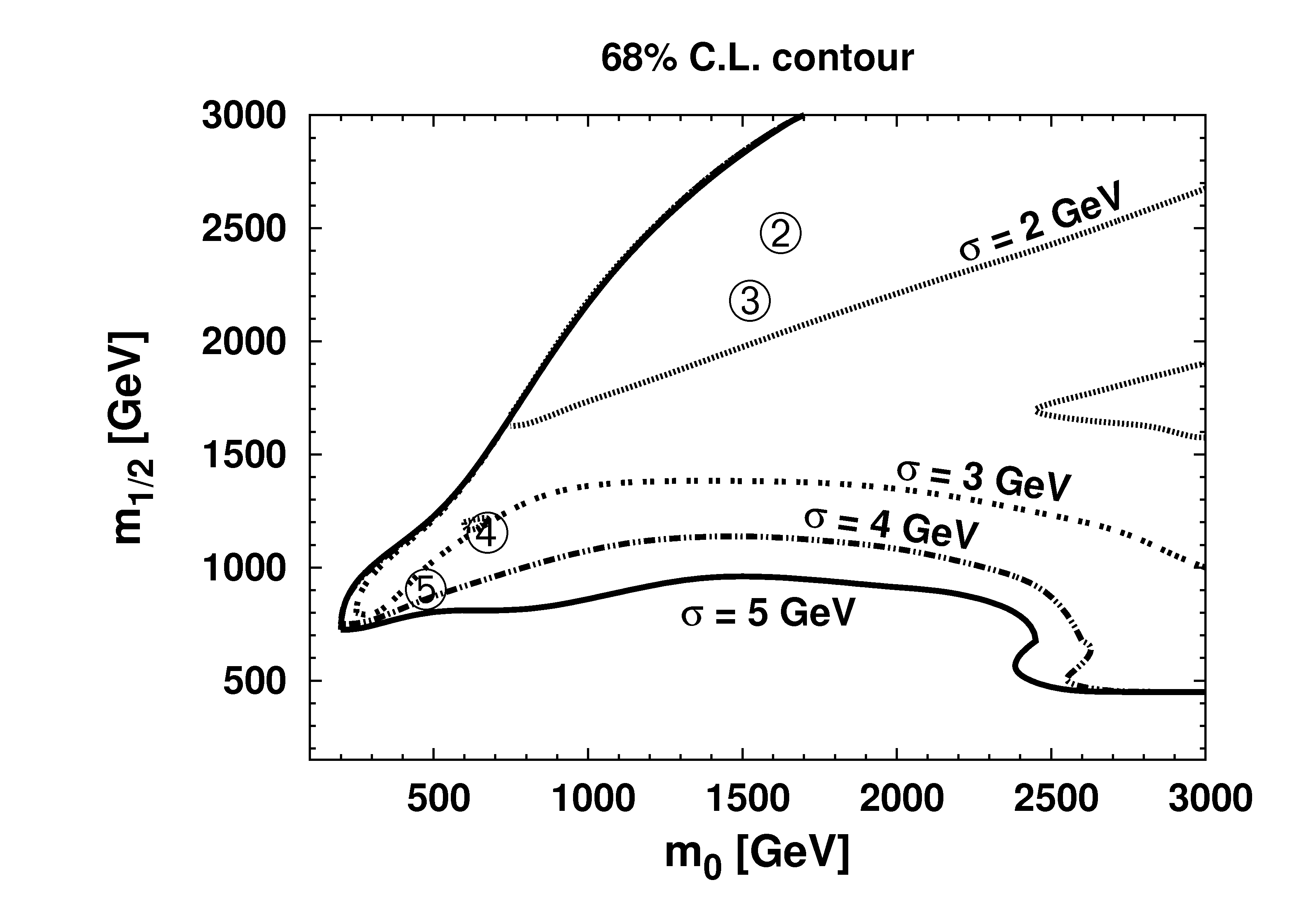}
 \includegraphics[width=0.49\textwidth]{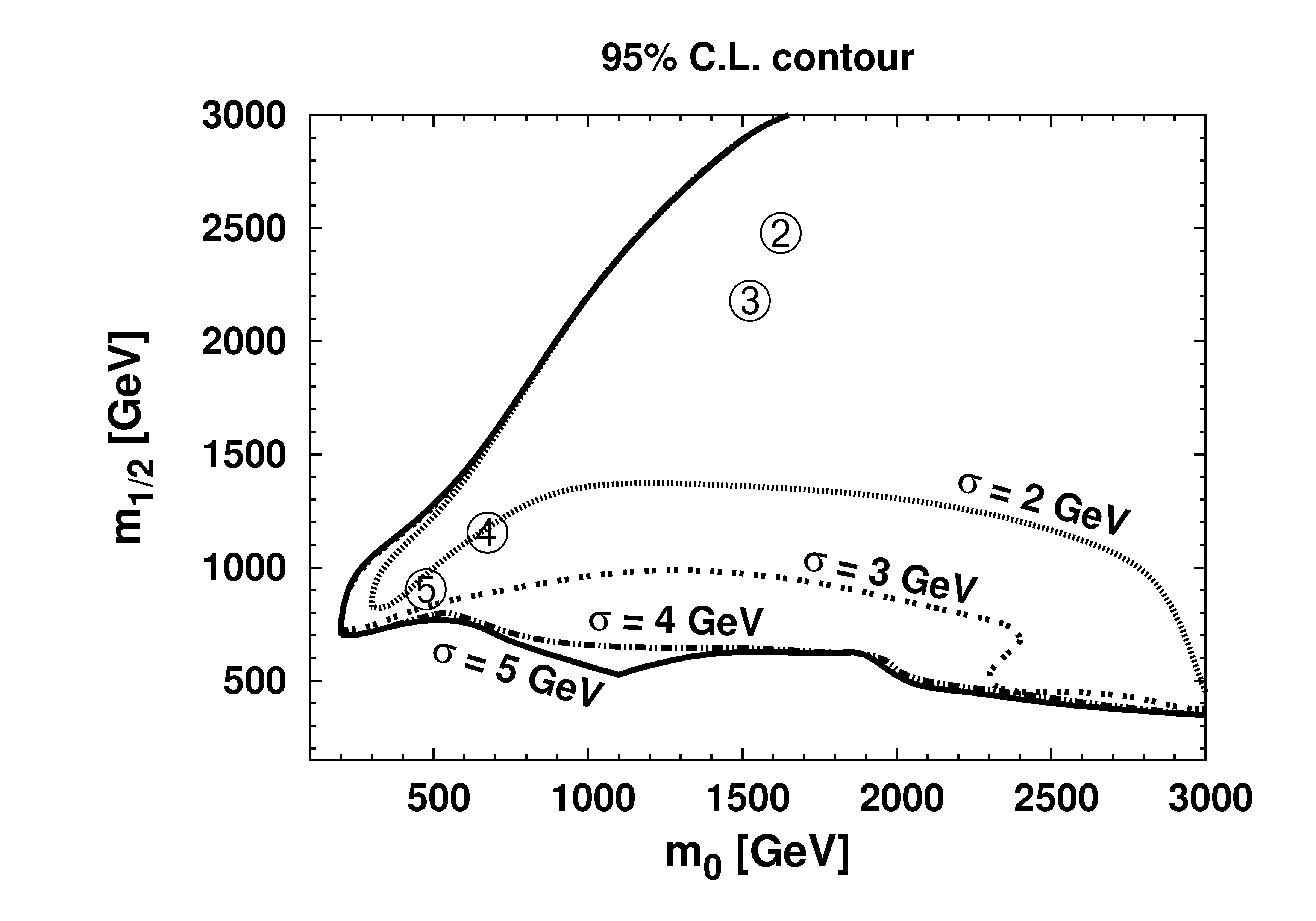}
 \end{center}
 \caption{If a Higgs mass of 125 GeV is imposed in the fit, the best-fit point moves to higher SUSY masses, but
 the location is strongly dependent on the assumed error for the  calculated Higgs mass. This error is indicated by the number inside the circle for the best-fit point. Left $\Delta\chi^2=2.3 (1\sigma)$ contour; right $\Delta\chi^2=5.99 (2\sigma)$ contour.  }
 \label{f5}
 \end{figure}

\subsection{Excluded region by direct searches for SUSY at the LHC}\label{lhc}

The direct searches for Supersymmetry at the LHC are dominated by the search for  strongly interacting particles, as shown in the publications by ATLAS \cite{ATLAS-CONF-2012-033} and CMS \cite{CMS-PAS-SUS-12-005}.   From  Fig.~\ref{f4} one observes that the excluded region (below the solid line) follows rather closely the total cross section for the production of squarks and gluinos $\sigma_{tot}$, indicated by the color shading. The 95\% C.L. on $\sigma_{tot} \left ( pp \rightarrow \tilde{g}\tilde{g},\tilde{g}\tilde{q},\tilde{q}\tilde{q} \right )$ is given by the contour line and varies between 0.003 and 0.03 pb, as shown in Table \ref{t1}. This is the motivation to approximate the $\chi^2$ contribution from the LHC experiments near the contour as $\sigma_{tot}^2/\sigma^2$, where $\sigma$ is defined by the requirement that at the exclusion limit the $\chi^2$ contribution equals 5.99 for each LHC experiment, as discussed in section \ref{multi}.    
  
  If the LHC data are combined with cosmological and electroweak data, the fitted values of \tb~ and the trilinear coupling vary, while the LHC experiments provided the exclusion contours for fixed values $\tb=10$ and $A_0=0$. 
  However, the efficiencies are not very sensitive to these parameters, so  this definition of $\chi^2_{LHC}$ has the advantage that the exclusion curve for each experiment is perfectly reproduced and that the dependence of the  cross section on $A_0$ and $\tb$ is taken into account. By adding the exclusion from each
 LHC experiment independently to the $\chi^2$ we assume tacitly that there are no correlations between these independent data sets.
 If we assume a 100\% correlation, we should only use a single experiment. However, the results would hardly change, as can be seen from a comparison from the combined curve in Fig. \ref{f1} and the published curves from each of the experiments: the difference in $m_{1/2}$ is at most 25 GeV.

\subsection{Excluded region by $\bsmm$}\label{bsmumu}
The upper limit on the branching ratio of $\bsmm$ can give significant constraints on the SUSY parameter space, since the $\bsmm$ rate varies as $tan^6\beta$. In addition $\bsmm$ is sensitive to the stop mixing which is a function of $A_0$. $\bsmm$ can be suppressed if $m_{\tilde{t_1}} \approx m_{\tilde{t_2}}$ or even get values below SM \cite{Beskidt:2011qf}. Hence  the $\chi^2$ is sensitive to the chosen $\tb$ and $A_0$ value. The combination with the relic density, which requires a large $\tb$ value in a large region of parameter space  (see Fig. \ref{f2})  causes  tension with the  $\bsmm$ constraint. This tension can be reduced by large values of $A_0$ as we showed in \cite{Beskidt:2011qf}, but with the recent upper limit near the SM value from LHCb \cite{Aaij:2012ac} this tension increased and both constraints cannot be fulfilled at the same time in the whole parameter space. This leads to two excluded
regions shown in Fig.~\ref{f1} by contour 2. The reason for the two regions is the following: at small values of $m_0$ the trilinear coupling $A_0$ cannot be made large enough to suppress $\bsmm$ enough, because then the stau leptons become tachyonic. At intermediate values
of $m_0$ the trilinear couplings can be made large enough to suppress $\bsmm$, but at larger values of $m_0$ $m_A$ becomes too large for large $A_0$ values, which leads to a  too large  relic density. Compared to the other constraints the excluded region by $\bsmm$ leads only to a tiny increase of the excluded region at small $m_0$.

\subsection{Effect of a SM Higgs mass $m_h$ around 125 GeV} 
The 95\% C.L. LEP limit of 114.4 GeV contributes for small and intermediate SUSY masses to the $\chi^2$ function, as shown by contour 3 in Fig. \ref{f1}. 
In the fit we use the  95\% C.L. LEP limit of 114.4 GeV on the Higgs mass instead of the limits published by CMS  and ATLAS with about 5/fb. 
In these publications CMS \cite{Chatrchyan:2012tx} and ATLAS \cite{ATLAS:2012ae} show some evidence for a Higgs with a mass around 125 GeV.  If we assume  this to be evidence for a  SM Higgs boson, which has similar properties as the lightest SUSY Higgs boson in the decoupling regime, we can check the consequences in the CMSSM we are investigating.  If a Higgs mass of 125 GeV is included in the fit, the best-fit point moves to higher SUSY masses, but there is a rather strong tension between the relic density constraint, $\bsmm$ and the Higgs mass, so the best-fit point depends strongly on the error assigned to the Higgs mass, as shown in Fig. \ref{f5}. The  experimental error on the Higgs mass is  about 2 GeV, but the theoretical error  can be easily 3 GeV. Therefore we have plotted the best-fit point for Higgs uncertainties between 2 and 5 GeV. One sees that the best-fit points wanders  by several TeV.  Clearly this needs a more detailed investigation in the future. It should be noted that the fit does not provide the maximum mixing scenario. If we exclude all other constraints, the maximum value of the Higgs can reach 125 GeV, albeit also at similarly large values of $\mhalf$. A negative sign of the mixing parameter $\mu$ shows similar results.

\subsection{Excluded region by the pseudo-scalar Higgs mass $m_A$}\label{relic}

The pseudo-scalar Higgs boson mass is determined by the relic density constraint, because the dominant neutralino annihilation contribution comes from A-boson exchange in the region outside the small co-annihilation regions. One expects $m_A \propto m_{1/2}$ from the relic density constraint, which can be fulfilled with \tb~ values around 50 in the whole ($\mzero,\mhalf$)-plane \cite{Beskidt:2010va}. Since the $m_A$ production cross section at the LHC is proportional to $\tan^2\beta$ the pseudo-scalar mass limit increases up to 496 GeV for the large values of $\tb$ preferred by the relic density (see Fig. \ref{f2}). 

In our fit we are not only using the relic density as a constraint but all data. The result of this optimization was shown in Fig.~\ref{f2} leading to different values of $A_0$ and $\tb$ in the ($\mzero,\mhalf$)-plane. The corresponding $m_A$-values are displayed in the right panel of Fig. \ref{f4} and the $m_A$ values excluded by the LHC searches lead to the excluded region, shown by the contour line in Fig. \ref{f4} (identical to contour 4 in Fig. \ref{f1} right).

\subsection{Excluded region by direct DM searches}\label{direct}

The cross section for direct scattering of WIMPS on nuclei has an experimental upper limit of about 10$^{-8}$ pb, i.e. many orders of magnitude below the annihilation cross section. This cross section is related to the annihilation cross section by similar Feynman diagrams. The many orders of magnitude are naturally explained in Supersymmetry by the fact that both cross sections are dominated by Higgs exchange and the fact that the Yukawa couplings to the valence quarks in the proton or neutron are negligible. Most of the scattering cross section comes from the heavier sea-quarks. However, the  density of these virtual quarks inside the nuclei is small, which is one of the reasons for the small elastic scattering cross section. In addition, the momentum transfer in elastic scattering is small, so the propagator leads to a cross section inversely proportional to the fourth power of the Higgs mass. 

Since the particle which mediates the scattering is typically much heavier than the momentum transfer, the scattering can be written in terms of an
effective coupling, which can be determined either phenomenologically from  $\pi N$ scattering or from lattice QCD calculations. 
The default values of the effective couplings in micrOMEGAs \cite{Belanger:2008sj} are: $f^{(p)}_{Tu}=0.033,~f^{(p)}_{Td} =
0.023,~f^{(p)}_{Ts}=0.26,~f^{(n)}_{Tu}=0.042,~f^{(n)}_{Td}=0.018,~f^{(n)}_{Ts}=0.26$.
The lower values from the  lattice calculations \cite{Cao:2010ph} are:
$f^{(p)}_{Tu}=0.020,~f^{(p)}_{Td} =
0.026,~f^{(p)}_{Ts}=0.02,~f^{(n)}_{Tu}=0.014,~f^{(n)}_{Td}=0.036,~f^{(n)}_{Ts}=0.02$.
Hence the most important coupling to the strange quarks vary from 0.26 to 0.02 \cite{Alarcon:2011zs},  which implies an order of magnitude  uncertainty in the elastic neutralino-nucleon scattering cross section.

Another normalization uncertainty in direct dark matter experiments arises from
the uncertainty in the local DM density, which
can take values between 0.3 and 1.3 GeV/cm$^3$, as determined from the
rotation curve of the Milky Way, see Ref. \cite{Weber:2009pt,deBoer:2010eh,Salucci:2010qr,Catena:2009mf} and references
therein.

  To get conservative estimates for the excluded regions, we take the
lowest possible values of the local DM density and the low couplings from lattice QCD calculations. 
The excluded region from the XENON100 cross section limit \cite{Aprile:2011hi} is shown by the contour line 5 in Fig. \ref{f1}. 
At large values of $\mzero$ EWSB forces the higgs\-ino component of the WIMP to increase and consequently the exchange via the  Higgs, which has an amplitude  proportional to the bino-higgsino mixing, starts to increase. This leads to an increase in the excluded region at large $\mzero$ and has here a similar sensitivity as the LHC.
 If we would take the less conservative effective couplings from the default values of micrOMEGAs the XENON100  limit would be 50\% higher than the LHC limit.

\section{Summary}

As mentioned in the introduction, several groups have performed similar analysis. Our results are closest to the one of Ref. \cite{Buchmueller:2011sw} if we adjust to the lower luminosity of 1/fb luminosity used in their analysis. However, in Ref. \cite{Buchmueller:2011sw} values of $\mhalf$ above 2500 GeV are excluded due to the tension with the relic density constraint \cite{heinemeyer}. In our case we do find good solutions and no such excluded region is found as shown in Fig.~\ref{f1}, left panel. This is probably due to the fact that in this region $\tb$ and $A_0$ are highly correlated, so they can be easily missed, if  SUSY samples are prepared with many, but not necessarily all, combinations of parameters. This region is the one preferred for Higgs masses around 125 GeV, as shown in Fig. \ref{f5}.

Our results differ significantly from results using Markov Chain Monte Carlo sampling. E.g. in Ref. \cite{Bertone:2011nj} values for intermediate values of $\mzero$ are excluded, which is the region of large $\tb$ (see Fig.~\ref{f2}, left panel). Here the parameters $\tb$ and $A_0$ are highly correlated and finding the correct minimum depends strongly on the stepping algorithm, e.g. stepping in the logarithm of a parameter is different from stepping in the parameter ("prior dependence"), see e.g. Ref. \cite{Trotta:2008bp}. Such dependence on sampling techniques largely disappears in our multistep fitting technique, since for each point of the ($\mzero,\mhalf$)-grid a unique solution is found independent of the minimizer used, so the frequentist approach with $\chi^2$ minimization yields the same results as a likelihood optimization with a Markov Chain sampling technique.

If one combines the limits from the direct searches at the  LHC, heavy flavor constraints, WMAP and XENON100 we exclude values of $\mhalf$ below  525 GeV in the CMSSM for $\mzero < 1500$ GeV, which implies a lower limit on the WIMP mass of 220 GeV and a gluino mass of 1270 GeV, respectively. For larger values of $m_0$ the excluded region drops to $\mhalf$ below 350 GeV, which leads to a lower limit on the LSP mass of 130 GeV and a gluino mass of 970 GeV, respectively. 

If a Higgs mass of the lightest Higgs boson of 125 GeV is imposed, the 
preferred region is well above this excluded region, but the size of the preferred region is strongly dependent on the size of the assumed theoretical uncertainty, as shown in Fig.  \ref{f5}. However, the higher than expected branching ratio and lower than expected branching ratio into tau leptons point to a Higgs with slightly different couplings from the SM. Such different couplings could exist in a supersymmetric model with an extended Higgs sector, like the NMSSM, see e.g. \cite{Ellwanger:2012ke}.

\section{Acknowledgements} 
Support from the Deutsche Forschungsgemeinschaft (DFG) via a Mercator Professorship (Prof. Kazakov) and the Graduiertenkolleg  "GRK 1694: Elementarteilchenphysik bei h\"ochster Energie und h\"ochster Pr\"azision"   is greatly appreciated. Furthermore, support from the Bundesministerium for Bildung und Forschung (BMBF) is acknowledged.

\section{Note added in proof} The CMS and ATLAS collaborations confirmed the published evidence for a new boson with a mass around 126 GeV at the beginning of the ICHEP 2012 conference \cite{pressrelease}, so the CMSSM would indeed prefer heavy SUSY masses, as shown in Fig. \ref{f5}. This conclusion does not change, if one relaxes the CMSSM constraints, e.g. by not requiring unification of the Higgs masses at the GUT scale. However, extended Higgs sectors would
need additional investigation.

\providecommand{\href}[2]{#2}\begingroup\raggedright\endgroup

\end{document}